\documentclass[]{article}

\pdfoutput=1
\usepackage{authblk}
\usepackage{fullpage}
\usepackage{amsmath}

\usepackage{caption}
\usepackage{graphicx}

\usepackage[hidelinks]{hyperref}

\usepackage[round, semicolon]{natbib}

\title{SWinvert: A workflow for performing rigorous surface wave inversions}
\author[1]{Joseph P. Vantassel}
\author[1]{Brady R. Cox}
\affil[1]{The University of Texas at Austin}

\begin{document}

\maketitle

\begin{abstract}
SWinvert is a workflow developed at The University of Texas at Austin for the inversion of surface wave dispersion data. SWinvert encourages analysts to investigate inversion uncertainty and non-uniqueness in shear wave velocity (Vs) by providing a systematic procedure and open-source tools for surface wave inversion. In particular, the workflow enables the use of multiple layering parameterizations to address the inversion's non-uniqueness, multiple global searches for each parameterization to address the inverse problem's non-linearity, and quantification of Vs uncertainty in the resulting profiles. To encourage its adoption, the SWinvert workflow is supported by an open-source Python package, \emph{SWprepost}, for surface wave inversion pre- and post-processing and an application on the DesignSafe-CyberInfracture, \emph{SWbatch}, that enlists high-performance computing for performing batch-style surface wave inversion through an intuitive and easy-to-use web interface. While the workflow uses the Dinver module of the popular open-source Geopsy software as its inversion engine, the principles presented can be readily extended to other inversion programs. To illustrate the effectiveness of the SWinvert workflow and to develop a set of benchmarks for use in future surface wave inversion studies, synthetic experimental dispersion data for 12 subsurface models of varying complexity are inverted. While the effects of inversion uncertainty and non-uniqueness are shown to be minimal for simple subsurface models characterized by broadband dispersion data, these effects cannot be ignored in the Vs profiles derived for more complex models with band-limited dispersion data. The SWinvert workflow is shown to provide a methodical procedure and a powerful set of tools for performing rigorous surface wave inversions and quantifying the uncertainty in the resulting Vs profiles.
\end{abstract}

\pagebreak

\section{Introduction and Literature Review}
Non-invasive surface wave testing has been increasingly recognized as a quick and inexpensive tool for measuring the elastic properties of the subsurface. The property of most interest is shear wave velocity (Vs), which is of critical importance in many areas of engineering and geophysics, including the applications of seismic site response \citep{teague_site_2016, vantassel_mapping_2018, teague_measured_2018, vantassel_multi-reference-depth_2019}, soil liquefaction prediction \citep{andrus_liquefaction_2000, wood_vs-based_2017}, and sub-surface imaging \citep{fathi_three-dimensional_2016, nguyen_site_2018}. Surface wave testing is performed in three stages: acquisition, processing, and inversion. Acquisition involves the measurement of an experimental wavefield along a site's surface. The experimental wavefield can be generated actively by those conducting the experiment (i.e., active-source) or the result of natural phenomena (i.e., passive-wavefield). A thorough summary of the best-practices for surface wave data acquisition can be found in the Guidelines for the Good Practice of Surface Wave Analysis \citep{foti_guidelines_2018}.

The second stage of surface wave testing is processing, and it involves the transformation of the experimental wavefield, which is predominantly composed of surface waves \citep{miller_partition_1955}, into experimental dispersion data. There are numerous ways to perform this wavefield transformation \citep{aki_space_1957, capon_high-resolution_1969, nolet_array_1976}, and while a full discussion is beyond the scope of this paper, they are, in general, dependent upon the acquisition setup (i.e., sensor layout and source location) and the wavefield type (i.e., active-source or passive-wavefield). Regardless of how the transformation is performed, the result of the processing stage is a measurement of the site's experimental dispersion relationship, which describes how the surface wave's phase velocity varies as a function of frequency ($f$) and/or wavelength ($\lambda$). The result of the processing stage is typically summarized into a single plot of frequency vs. surface wave phase velocity, which is commonly and hereafter referred to as the experimental dispersion data. Since the acquisition and processing stages contain some level of uncertainty, attempts should be made to quantify the dispersion uncertainty \citep{lai_propagation_2005, foti_non-uniqueness_2009, cox_surface_2011, teague_development_2018}.This quantified uncertainty should be propagated forward into the inversion stage by representing the phase velocity at each frequency with a mean and standard deviation. This practice is consistent with the belief that the uncertainty in experimental dispersion data is normally distributed \citep{lai_propagation_2005}.
 
\begin{figure}[t]
\includegraphics[width=\textwidth]{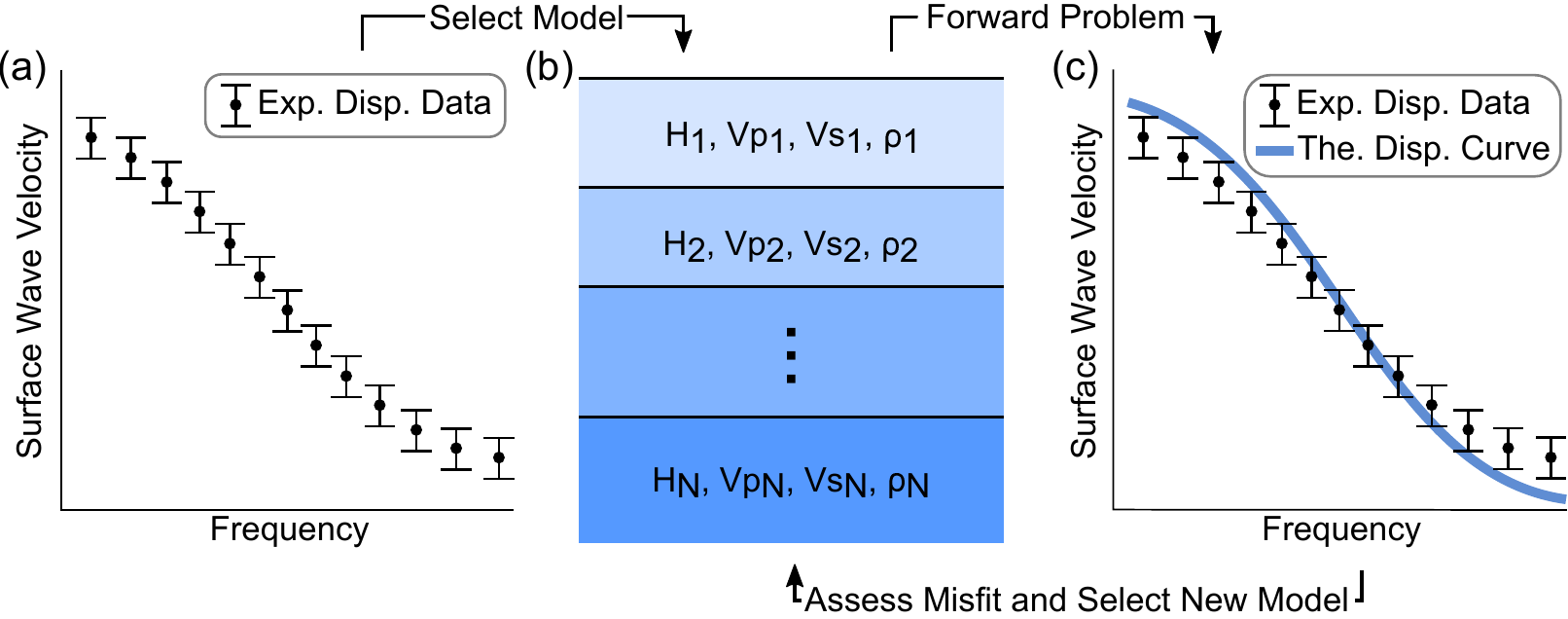}
\caption{Schematic representation of the inversion process, beginning with: (a) the development of experimental dispersion data with measures of uncertainty, (b) selection of a candidate layered earth model with specific geometric and elastic properties, and (c) comparison of the candidate's theoretical dispersion curve with the experimental dispersion data, assessment of the misfit, and selection of a new candidate layered earth model.}
\label{fig:schinv}
\end{figure}

The third stage of surface wave testing is inversion, and it involves finding the velocity model(s) whose theoretical dispersion curve(s) generated through solution of a forward problem \citep{thomson_transmission_1950, haskell_dispersion_1953, gilbert_propagation_1966, kausel_explicit_1981} match(es) the experimental dispersion data from the processing stage. As there is no unique way to transform experimental dispersion data directly into a subsurface velocity model, numerical search techniques are required. These techniques involve searching large numbers of potential subsurface velocity models to find those whose theoretical dispersion curve best matches the experimental dispersion data. The goodness of fit between the theoretical and experimental dispersion is assessed through the use of a quantitative misfit value/function, which is generally a L2 norm of residuals or a normalized version thereof [e.g., root-mean-square error (RMS)]. Figure \ref{fig:schinv}, illustrates schematically the inversion of experimental dispersion data with uncertainty. Figure \ref{fig:schinv}a shows the experimental dispersion data extracted from the processing stage. Figure \ref{fig:schinv}b shows a candidate/trial layered earth model. Figure \ref{fig:schinv}c shows the candidate model's theoretical dispersion curve, computed via the forward problem, in comparison to the experimental dispersion data. The iterative nature of the inverse problem is indicated by the calculation of the model's misfit and selection of a new candidate model, as shown at the bottom of Figure \ref{fig:schinv}c. Inversion should be considered the most challenging part of surface wave testing and the practical considerations of its performance are the primary focus of this work.

The challenges associated with the surface wave inverse problem have been discussed widely in the literature and include the inverse problem's significant non-linearity, mixed-determined nature, and ill-posedness \citep{foti_surface_2015}. The significant non-linearity of the problem indicates that the experimental dispersion data and the parameters of a given layered velocity model are not linearly related. Therefore, tools for the solution of linear problems, as are readily available in many branches of mathematics such as linear algebra, may not be used without serious risk of oversimplification. The problem's mixed-determined nature means the solution can be both over-determined (more-equations than unknowns) and under-determined (more-unknowns than equations) at the same time. This is a result of the non-invasive nature of surface wave testing. Since surface wave testing is, as the name implies, performed at the surface, the properties at shallow depths may be well-constrained, whereas those at greater depths are substantially less so, leading to the problem's mixed-determined nature. And finally, the problem is ill-posed, meaning a unique solution is not guaranteed as the problem is unstable \citep{hadamard_lectures_1923}, and so it is possible for many different velocity models to fit the measured experimental dispersion data equally well. A successful inversion workflow must demonstrate its ability to address each of these complicating factors.

Of paramount importance to the workflow is the surface wave inversion algorithm itself. The literature contains many different types of search algorithms, which can be divided broadly into two categories: local and global. Local search algorithms, also known as linearized inversion methods, rely on an initial starting model, which is assumed to be close to the true solution. This assumption may or may not be valid and is unverifiable for real data. Using the starting model, the sensitivity of each model parameter is calculated assuming a linear relationship (i.e., 1st order Taylor series expansion) between the parameter of interest and the model's theoretical dispersion misfit. Using the sensitivity of each parameter a new model is selected that is predicted to decrease the solution's misfit. This process is repeated until some desired misfit, change in misfit, or maximum number of allowed iterations is reached. Most commercial surface wave inversion software packages use this approach due to its computational speed and ease of use. 

In contrast, global search algorithms rely upon upper and lower limits of each model parameter, which as a whole is often referred to as the parameterization space. Unlike local search methods, global search methods do not require the user to provide a single initial starting model, and instead the process of selecting candidate models that fall within the parameterization space is left to the algorithm. Global search algorithms include grid-based searches, Monte Carlo \citep{sambridge_monte_2002, socco_improved_2008}, simulated annealing \citep{sen_nonlinear_1991}, genetic algorithms \citep{lomax_finding_1994}, importance sampling methods \citep{sen_bayesian_1996}, and neural networks \citep{meier_initial_1993}. The most common are the grid-based, Monte Carlo, and importance sampling methods, of which importance sampling methods are generally preferred as they are able to focus on the most promising portions of the parameter space and thereby offer significant computational gains over grid-based searches.

When using global searches, it is of paramount importance that the parameter space be sufficiently large to ensure it contains the true solution, while not overly permissive so as to further confound the search for the true solution. It is important to note that researchers have developed global search algorithms that can search outside of the initial parameter space \citep{socco_improved_2008}, as well as algorithms that use a global search for selecting initial models for subsequent refinement by a local search \citep{caylak_inversion_2012}. While these search algorithms show promise, they have not been released openly, making it impractical to utilize them in an open-source inversion workflow. Instead, the algorithm utilized in the proposed workflow is an importance sampling method developed by Sambridge \citep{sambridge_geophysical_1999, sambridge_geophysical_1999-1} called the Neighborhood Algorithm (NA). The NA has since been implemented in the Dinver surface wave inversion module of the open-source Geopsy software by Wathelet \citep{wathelet_surface-wave_2004, wathelet_array_2005, wathelet_improved_2008}. Dinver and the NA have become quite popular for the inversion of surface wave data, thereby making it an ideal inversion engine to be adopted into the SWinvert workflow.

This paper details a systematic workflow for inverting experimental dispersion data to obtain layered velocity models. The workflow focuses on making specific recommendations regarding the three main stages of the inversion process: (1) development of the inversion's target, (2) development of the inversion's parameterization, and (3) selection of the inversion's tuning parameters. These recommendations are then validated using twelve synthetic models of varying complexity to demonstrate its robustness and limitations. Following the inversion of the twelve synthetic models, two of the twelve datasets, one termed a high-variance and the other a low-variance dataset, are selected to demonstrate an approximate approach to quantifying Vs uncertainty in surface wave inversion. The uncertainty in both the high-variance and low-variance case is shown to be less than what is typically assumed in practice. Finally, specific guidance and recommendations to those wishing to use this workflow are summarized.

\section{Development of the Inversion's Target}	
The SWinvert workflow begins at the end of the processing stage, where the user has developed experimental dispersion data for their site. Preferably, this experimental dispersion data would include site-specific estimates of uncertainty at each frequency/wavelength. However, if site-specific dispersion uncertainty has not been quantified, typical values for dispersion uncertainty documented in surface wave blind studies published in the literature \citep{cox_synthesis_2014, garofalo_interpacific_2016} can easily be associated with the dispersion data using the Python package \emph{SWprepost} \citep{vantassel_jpvantasselswprepost_2020} associated with SWinvert. To provide a specific recommendation that is consistent with those references previously mentioned, a surface wave phase velocity coefficient of variation between 0.05 and 0.10 is recommended for use at most sites where site-specific uncertainty has not been quantified.

\subsection{Dispersion Curve Resampling}
The first stage of the inversion process is developing the target for the inversion algorithm. For simplicity in explaining the workflow, this paper will focus primarily on fundamental-mode Rayleigh wave experimental dispersion data, although this is not the only type of inversion target (discussed later). Of most relevance to the development of an inversion target is selecting how the experimental dispersion data is to be represented, as this can significantly affect the inversion's computation time and solutions it produces.
As most processing methods for active-source and passive-wavefield data utilize the Fast-Fourier Transform to convert time-domain wavefield recordings to the frequency-domain, the experimental dispersion data from many software packages is provided in terms of linear-frequency. While commonly performed, presenting dispersion data in terms of linear frequency is not ideal for the inversion process. Consider, for example, dispersion data that is sampled in linear-frequency between 1 and 50 Hz with a frequency step of 1 Hz, resulting in a total of 50 data points. Of these points, half would reside above 25 Hz, which, assuming a typical Rayleigh wave phase velocity (Vr) for soft ground of 150 m/s, would result in wavelengths less than 6 m. As the penetration depth of a Rayleigh wave can be approximated by its wavelength divided by a depth factor of 2 or 3, this would result in a majority (\textgreater50\%) of the data characterizing the upper 2 m to 3 m, whereas the other half could be responsible for characterizing down to a depth of 50+ m, depending on how the profile's stiffness increases with depth. This example illustrates that linear-frequency sampling should not be used, as it will bias the results toward predominantly fitting the near-surface structure.

To overcome the disadvantages of linear-frequency sampling, the common practice is to resample the dispersion data in log-frequency. While this is a significant improvement over linear-sampling, it also has one minor disadvantage. Namely, for those dispersion curves where Vr changes rapidly in terms of frequency (e.g., due to a significant impedance contrast in the velocity profile), resampling in terms of log-frequency can result in what appears to be gaps in the dispersion data, where the trend of the data cannot be well represented without using additional samples. The use of additional samples incurs additional computational cost in the inversion, which is undesirable. As an alternative, we recommend resampling dispersion data in terms of log-wavelength. Vr as a function of wavelength tends to change less rapidly point-to-point than in terms of frequency, thereby mitigating spurious gaps in the dispersion data without incurring the additional computational cost of increasing the number of samples. Additionally, viewing dispersion data in terms of wavelength (refer to Figure \ref{fig:param}a) is very helpful in determining appropriate inversion parameters, such as the minimum layer thickness, which is a function of the minimum resolved wavelength ($\lambda_{min}$), and maximum depth of investigation, which is a function of the maximum resolved wavelength ($\lambda_{max}$), as discussed in more detail below. In summary, we recommend resampling the experimental dispersion data in log-wavelength with a sufficient number of points to capture the general trends/kinks/bends in the data (typically a minimum of 20 -- 30). The supporting Python package, \emph{SWprepost}, includes the functionality to resample experimental dispersion data in various ways, including the recommended method of log-wavelength. \emph{SWprepost} allows for writing resampled dispersion data to a basic text file or the .target format used by Dinver.

\subsection{Utilization of Additional Target Information}
While the main focus of this work is the inversion of fundamental mode Rayleigh-type surface wave dispersion data, joint inversions that include additional sources of information deserve a brief discussion. Researchers have used additional target information for inversion including: higher-mode Rayleigh wave dispersion data along with fundamental- and higher-mode Love wave dispersion data \citep{yust_epistemic_2018}, and the fundamental frequency of the horizontal-to-vertical spectral ratio to constrain the Rayleigh wave ellipticity \citep{vantassel_mapping_2018, yust_epistemic_2018, teague_measured_2018}. These joint inversions, when done properly, have been shown to reduce the uncertainty in Vs profiles derived from surface wave measurements by providing additional information to mitigate the problem's non-uniqueness and ill-posedness. However, the utilization of additional target information should be done carefully because constraining the solution with erroneous/poor-quality data may bias the final result while simultaneously reducing the apparent uncertainty, making accurate predictions about the true profile difficult, if not impossible. While the initial version of \emph{SWprepost} released in conjunction with this paper focuses on fundamental-mode Rayleigh dispersion data, all of the additional constraints aforementioned (higher-mode Rayleigh, fundamental- and higher-mode Love, and Rayleigh wave ellipticity) can be readily pre-processed using the package and then appended to the fundamental-mode Rayleigh data using the Dinver interface.

\section{Development of the Inversion's Parameterization}
The second step of the inversion workflow is the definition of the parameter space in which each candidate layered earth model must reside in order to be considered. Generally, the parameter space is discretized into an assumed number of layers (refer to Figure \ref{fig:schinv}b). Each layer must be assigned a range, or fixed value, of thickness (H), mass density ($\rho$), shear stiffness (typically Vs), and compression stiffness (typically Vp). This layer-discretized form of the parameter space is generally and hereafter referred to as the inversion's parameterization. The development of a high-quality parameterization can be difficult and time-consuming, as the appropriateness (or inappropriateness) of a trial parameterization may not be clear until after an inversion has been run. Therefore, to encourage high quality parameterizations, additional site information should be used whenever it is available \citep{teague_development_2015, teague_development_2018, cox_layering_2016}. This a priori information may include general geologic information, preliminary P- and S-wave velocity models from P- and S-wave refraction, and invasive geotechnical testing such as the standard penetration test (SPT) and cone penetration test (CPT). These methods will lend general information on the site's near-surface layering and stiffness, though caution should be taken against overly constraining the inversion's parameterization based on any one source of information, as each has its own unique limitations. Consider, for example, constraining surface wave inversions using results from a non-invasive test such as P-wave refraction and an invasive test such as a CPT. Refraction will tend to produce subsurface models that are an average of the material beneath the lateral extent of the array, while CPT will provide a very detailed measurement of the layering beneath a single point on the surface. While both types of testing provide useful information that should be used to constrain the inversion, neither should be accepted with complete certainty in terms of parameterizing the surface wave inversion. The parameterization should be restricted to prevent the occurrence of unrealistic velocity models, but left sufficiently permissive to not over-constrain the problem and bias the results.

As surface wave methods are being increasingly used and marketed as a preliminary form of site characterization, a priori information may not be available. Or, if site characterization data is available it may not extend to sufficient depth to provide information on the site layering and stiffness where needed. In either case, there is a need for a systematic procedure that can be used to develop several potential parameterizations based on the theoretical dispersion data alone. Previous researchers have studied the impact of using different parameterizations to invert surface wave dispersion data using a synthetic blind study and field case history \citep{cox_layering_2016} and large field data sets \citep{di_giulio_exploring_2012, hollender_characterization_2018}.  However, to the author's knowledge, no one has performed a large scale synthetic study to examine the effect of using multiple parameterizations, which is one of the primary focuses of this work. In the absence of a priori information, the development of a parametrization can be divided into two general stages. The first stage is the selection of a number of layers and their potential thicknesses (or bottom depths). The second stage is the selection of a range of values (Vs, Vp, etc.) for those layers. This study will focus primarily on the former, and assume broad ranges for the latter to consider the worst case where little information beyond the fundamental mode experimental dispersion data is known. 

\begin{figure}[t]
	\includegraphics[width=\textwidth]{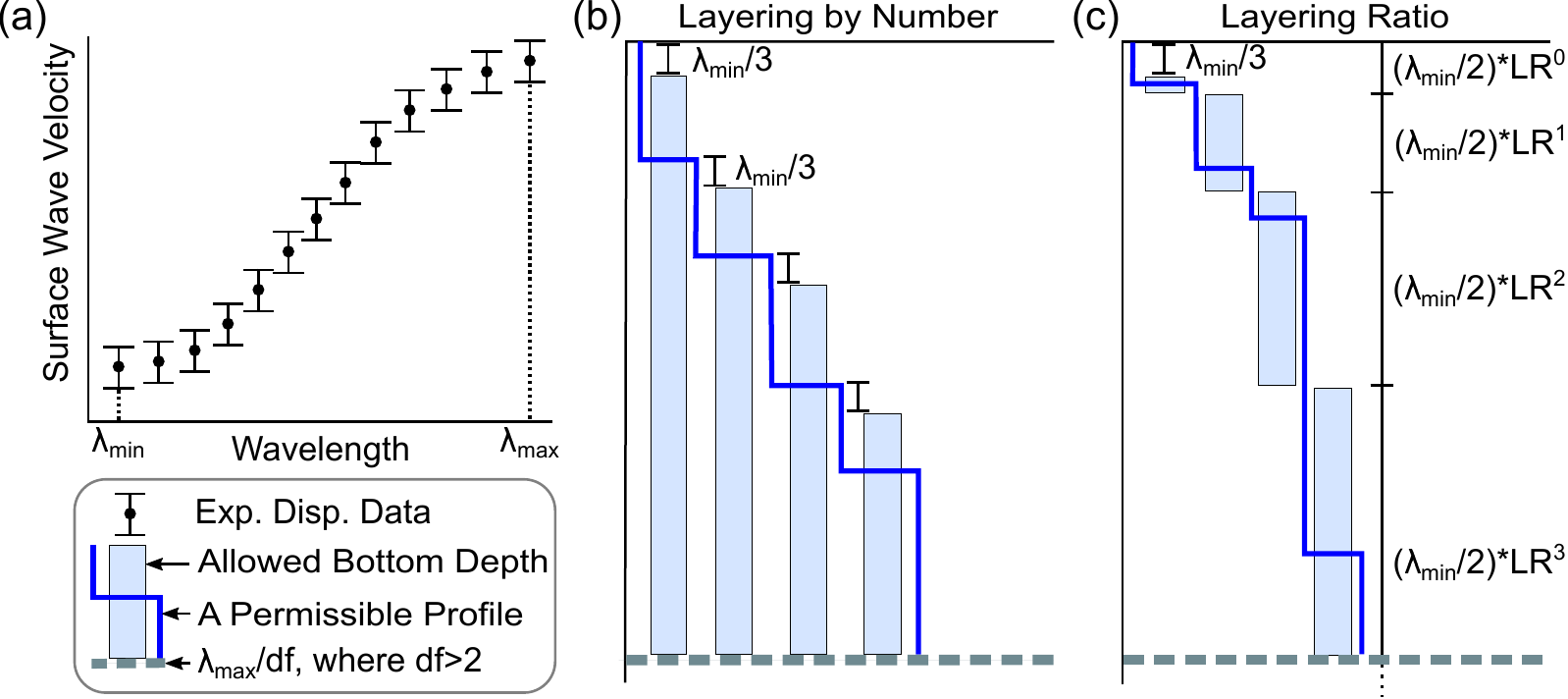}
	\caption{(a) experimental dispersion data in terms of wavelength, (b) schematic representation of a layering by number (LN) inversion parameterization equal to 4, and (c) schematic representation of a layering ratio (LR) inversion parameterization equal to 2.0. Both parameterizations are fully defined by the experimental dispersion data and only two user-specified parameters: the number of trial layers (in the case of LN) or the layering ratio (in the case of LR) and the depth factor (df), which is typically either 2 or 3. Both parameterizations utilize the minimum and maximum measured wavelength ($\lambda_{min}$and $\lambda_{max}$, respectively) of the fundamental mode experimental dispersion data to determine appropriate minimum layer thicknesses and the maximum allowable profile depth.}
	\label{fig:param}
\end{figure}

This study will examine the effect of developing parameterizations using two potential layering schemes. The first layering parameterization scheme is Layering by Number (LN). A schematic example for a Layering by Number parameterization of 4 (LN=4) is shown in Figure \ref{fig:param}b. The selection of LN=4 determines that the parameterization will have 4 layers, including the bottom-most half-space. The minimum thickness of any layer is defined by the minimum measured dispersion data wavelength divided by 3 (i.e., $\lambda_{min}/3$), and the maximum bottom-depth of any layer is defined by the maximum measured wavelength divided by a user defined depth factor (i.e., $\lambda_{max}/df$), where df is typically 2 or 3 \citep{garofalo_interpacific_2016, foti_guidelines_2018}. The inversion algorithm is free to explore models with any potential layer thicknesses that exist within these broad allowable bottom-depth ranges. Note that since each layer has a minimum required thickness, the minimum allowed bottom-depth of any layer is dependent solely on the bottom of the layer above it (selected by the inversion algorithm) and $\lambda_{min}$. Whenever possible, a LN parameterization number should be chosen to match the known, or suspected, number of subsurface layers at the site. If the number of subsurface layers is unknown, as is most often the case, the analyst will need to try several different LN parameterizations as a means to investigate inversion uncertainty and non-uniqueness. The SWinvert workflow and its supporting software tools offer the ability to quickly and easily consider multiple LN parameterizations. Typical values for LN range between 3 and 15. Note that we do not recommend blindly using a large number of thin layers with uniform thickness in the hopes of more accurately resolving subsurface variations, as this often results in profiles that have abrupt and unrealistic changes in Vs. 
 
The second inversion parameterization method is the Layering Ratio (LR) approach proposed by \citet{cox_layering_2016}. A schematic example for a Layering Ratio of 2.0 (LR=2.0) is shown in Figure \ref{fig:param}c. Starting from the first layer, the bottom-depth must exist within the range of the minimum wavelength divided by 3 and 2 (i.e., $\lambda_{min}/3$ and $\lambda_{min}/2$, respectively). The range of potential bottom-depths for each subsequent layer is defined by the product of the maximum possible bottom-depth of the first layer (i.e., $\lambda_{min}/2$) and LR raised to the layer's number, where the layers are numbered from zero rather than one. There are two details of note that should be discussed in regards to the LR method. First, the LR used in this workflow is a slight modification of the original proposed by \citet{cox_layering_2016}. In their paper, \citet{cox_layering_2016} recommended that no layer should begin below the maximum resolved depth of profiling (i.e., $\lambda_{min}/df$), as illustrated in Figure \ref{fig:param}c. However, this recommendation was not formalized into the equations governing the setup of the LR and was left to the judgement and vigilance of the analyst. In the modified version of the LR approach presented here, the recommendations of \citet{cox_layering_2016} are formalized to ensure that no layers begin below the resolution depth. Second, while the LR method permits thinner layers beneath thicker layers, their use requires the layer beneath the thin layer to be relatively thick. As the example profile in Figure \ref{fig:param}c shows, the inclusion of a thin layer (layer 2, recall the layers are numbered from 0) beneath a thicker layer is permitted, although the following layer (layer 3) is ``punished" for the presence of the overlying thin layer because its bottom depth must now extend down through the majority of the bottom-depth range of layer 2 and into its own potential bottom-depth range. This effect can be mitigated by using smaller LR values, noting that a smaller LR will result in a larger numbers of layers in the parameterization that tend to be thinner. As noted above in regards to LN parameterizations, the analyst will need to try several different LR parameterizations as a means to investigate inversion uncertainty. As with LN, the SWinvert workflow and its supporting software tools offer the ability to quickly and easily consider multiple LR parameterizations. Typical values for LR include 1.2, 1.5, 2.0, 3.0, 3.5, 5.0 and 7.0.

Both the LN and LR parameterizations have been programmed into \emph{SWprepost} such that each parameter (i.e., Vp, Poisson's ratio, Vs, and mass density) may be defined with the same or different layering schemes, allowing the user to mix-and-match as desired. Note that Poisson's ratio is not an independent parameter of the ground model (as it is uniquely defined by the values of Vp and Vs), and in Dinver it serves solely as a check that the selected values of Vp and Vs are consistent with the expected range of Poisson's ratio for that layer. As a velocity model's dispersion data is most sensitive to Vs, much less so to Vp, and negligibly so to mass density \citep{wathelet_array_2005}, one common parameterization strategy aimed at reducing computation time is to use the largest number of layers for Vs, fewer layers for Vp, and only a few or a single layer for mass density. For example, if a site was known to be composed of varied soil layers over rock, a reasonable parameterization would include: a two-layer mass density model (one for soil and one for rock), a simple parameterization for Vp and Poisson's ratio, such as LN=3 to account for three major Vp layers (e.g., dry soil, saturated soil, and rock), and the utilization of multiple LN and/or LR parameterizations with different numbers of layers for Vs (discussed at length below) to search for the best subsurface layering model(s) and account for uncertainty in the primary parameter of interest. Once the type of layering and appropriate ranges for each parameter have been selected, \emph{SWprepost} can be used to setup the parameterization files and export them to the .param format used by Dinver, removing the time-consuming and error-prone step of entering them manually.

\section{Selection of the Inversion's Tuning Parameters}
Following the selection of the inversion's targets and parameterizations, the tuning parameters for the inversion algorithm need to be selected. For versions of Geopsy prior to 3.0.0 there are four tuning parameters of significance: the number of initial models (Ns0), number of NA iterations (It), number of models per iteration (Ns), and the number of models to consider when resampling (Nr). To provide some physical insight into the effects of these parameters, we briefly summarize how the NA works with respect to each of these parameters. First, Ns0 models are randomly sampled from the parameter space, theoretical dispersion curves are calculated, and misfits relative to the experimental data are calculated. Second, at most Nr of these ``best" (i.e., lowest misfit) models are used to define regions of interest (neighborhoods) in which the inversion is going to focus during the next iteration. Third, the first iteration begins with the selection of Ns new models based on the neighborhoods discovered in the previous iteration. This process continues with the selection of new neighborhoods and Ns new samples until all It iterations have been completed. From this brief summary, it follows that the total number of models searched is the sum of Ns0 and the product of It and Ns (i.e., Ns0 + It*Ns). As part of a large parametric study, described in detail below, the relative values of It and Ns (e.g., It large, Ns small; It medium, Ns medium; It small, Ns large) were found to be insignificant compared to the value of their product (i.e., It*Ns). We therefore decided to combine them into a single tuning parameter representing the total number of models searched with the NA. Of note to the reader, for versions of Geopsy including and after 3.0.0, which was released during this study, there are only three tuning parameters of significance: the number of initial models (Ns0), number of models searched by the NA (Ns), and the number of models considered during resampling (Nr), where Ns has been redefined as the product of It and the former Ns. This thereby independently corroborates our decision to combine Ns and It into a single representative tuning parameter It*Ns. 

\begin{figure}[t]
	\includegraphics[width=\textwidth]{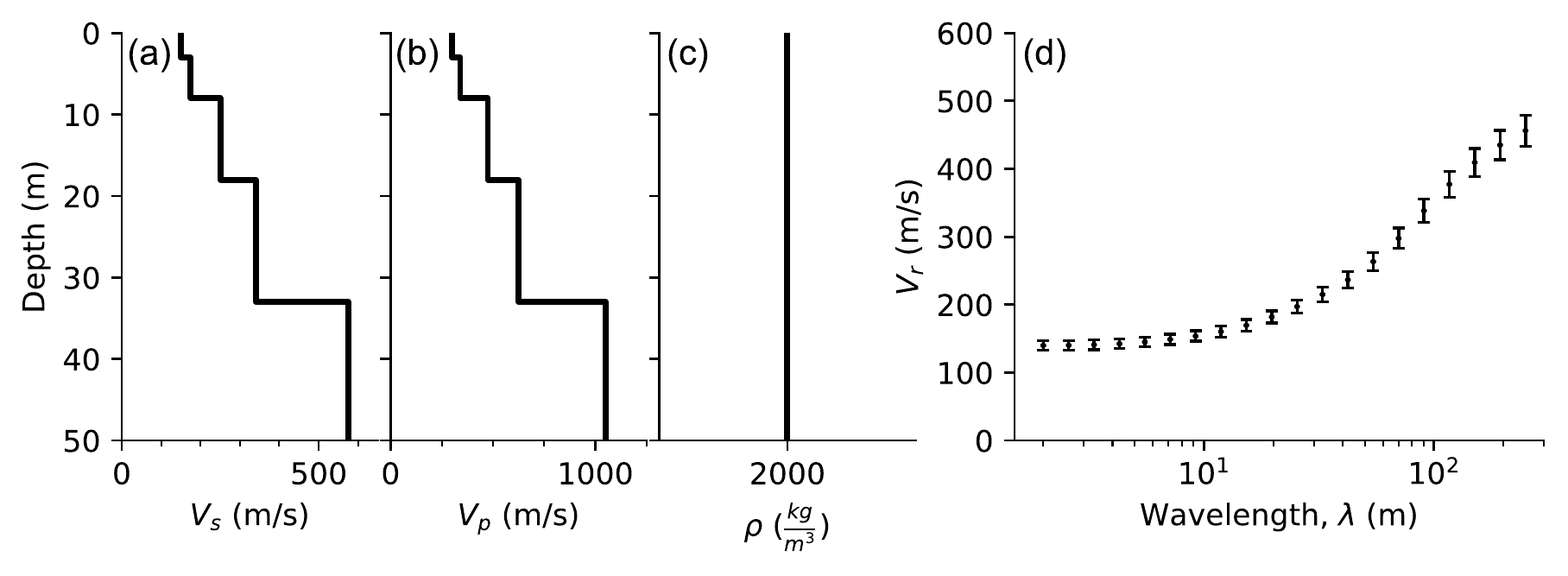}
	\caption{Velocity model and experimental dispersion data used to investigate Dinver's tuning parameters. The model is described in terms of its: (a) shear wave velocity (Vs), (b) compression wave velocity (Vp), (c) mass-density ($\rho$), and (d) experimental dispersion data with assumed 5\% coefficient of variation (COV) on Rayleigh phase velocity (Vr) to represent dispersion uncertainty.}
	\label{fig:tunemodel}
\end{figure}

To determine what values should be assigned to these tuning parameters, a large parametric study was performed. The parametric study involved the repeated inversion of synthetic experimental dispersion data derived from a prescribed velocity model containing five layers of varying Vs and Vp and a single, constant mass density layer (refer to Figures 3a, b, and c). The mean synthetic dispersion data for this model was computed using the Thomson-Haskell forward problem, resampled in log-wavelength using 20 samples between 2 and 250 m, and a standard deviation consistent with a 0.05 coefficient of variation (refer to Figure \ref{fig:tunemodel}d). This synthetic dispersion data was then inverted as a means to investigate the inversion's sensitivity to variable tuning parameters. As it was possible that the two parameterization formats previously discussed (i.e., LN and LR) could perform differently, the study was completed using both a LN=5 and LR=2.5 Vs parameterization. LN=5 was selected because it contained the same number of layers as the true Vs profile and layer boundaries that permitted a match to the true Vs profile. Meeting these conditions was unfortunately not possible for the LR approach. Therefore, LR=2.5 was selected so that its layer boundaries permitted a match to the true Vs profile, although it did include one additional near-surface layer (i.e., six total layers instead of five). For this part of the study, the parameterizations of Vp and mass density were kept fixed at the true values. An important point to note here is that this scenario should be considered the ``best-case" for surface wave inversion, where the mean of the dispersion data includes no error, Vp and mass density are known, the parameterizations for Vs are consistent with the true number of layers, and the limits on Vs (100 -- 700 m/s) encompassed the true solution. Thus, the insights from this study as to how many models need be explored during inversion should be considered a lower-bound when applied in practice.

The values considered for each tuning parameter include Ns0 = 5, 50, 500, and 5000; Nr = 5, 50, 100, and 500; and both It and Ns = 50, 100, 500, 1000, and 5000. To limit the computational expense of performing these inversions, no combination where the total number of models searched with the NA (It*Ns) exceeded 250,000 were considered. As will be discussed in detail, inversions performed using the same parameterization and the same number of models resulted in ground models of variable quality. This variability is believed to be due to the stochastic nature of the inversion algorithm and the problem's inherent non-linearity. Regardless of the cause, each tuning parameter combination was repeated 10 times, as independent Dinver inversion runs, to examine the variability and ensure a reliable estimate of convergence. In total, between the two LN and LR parameterizations, nearly 400 million models were searched in this tuning parameter study.
 
\begin{figure}[t]
	\includegraphics[width=\textwidth]{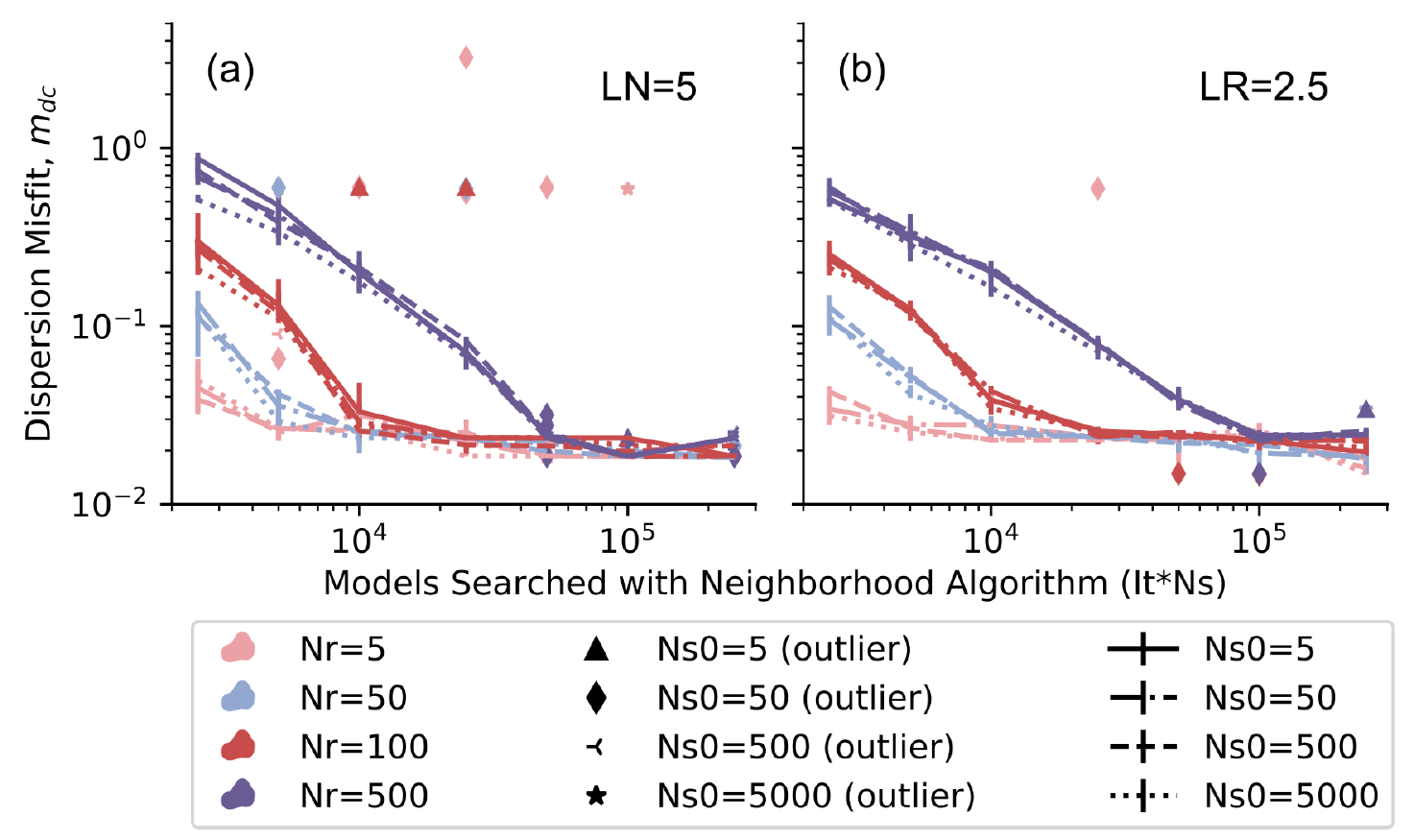}
	\caption{Effect of inversion tuning parameters Nr, Ns0, and It*Ns on the minimum dispersion misfit ($m_{dc}$) when using a shear-wave velocity parameterization of: (a) Layering by Number of 5 (LN=5) and (b) Layering Ratio of 2.5 (LR=2.5). Color denote the value of Nr, line style the median value of Ns0, vertical bars the interquartile range of $m_{dc}$ for 10 independent trials, and marker type the value of Ns0 for $m_{dc}$ outliers (i.e., those outside 5 times the interquartile range).}
	\label{fig:tunestudy}
\end{figure}

The effects of each tuning parameter on the minimum dispersion misfit ($m_{dc}$) achieved during each trial inversion is illustrated in Figure \ref{fig:tunestudy}. Note that the misfit function used herein is the normalized root-mean-square error proposed by \citet{wathelet_surface-wave_2004}. The primary variables of interest are represented in Figure \ref{fig:tunestudy} in the following ways: It*Ns by the abscissa, $m_{dc}$  by the ordinate, Nr by color, median Ns0 by line style, the $m_{dc}$  interquartile range by the vertical bars, and results outside 5 times the interquartile range of $m_{dc}$  by marker type. The effect of and interaction between these tuning parameters will be discussed in turn. First, in the order from most to least significant, is It*Ns, whose value tends to be inversely correlated with the minimum $m_{dc}$ , indicating that large values of It*Ns should result in smaller $m_{dc}$  (i.e., better solutions). However, the inverse relationship does not hold for all values of It*Ns, and so at some point increasing It*Ns tends to no longer have an effect on the minimum $m_{dc}$  achieved. This indicates that running an inversion with more models does not guarantee a better (i.e., lower misfit) result when a smaller but sufficient number of models has already been considered. Second, Nr is observed to have a slowing effect on the inversion's convergence, with lower values of Nr tending to converge faster (i.e., after fewer models) than higher values. However, this faster convergence is juxtaposed by the tendency of low Nr inversions to become trapped in local minima and, as a result, increase the number of $m_{dc}$  outliers (i.e., high $m_{dc}$  values), even when a large number of models has been considered (e.g., It*Ns=10\textsuperscript{5} and Nr=5 in Figure \ref{fig:tunestudy}a). Thus, there is a trade-off between fast-convergence to a low misfit solution and a full and rigorous exploration of the model space. The danger in using a low Nr value is that some inversions may not converge to a good model. This effect could not have been observed without running multiple distinct trials. Finally, Ns0 is shown to have a minimal effect on the misfit attained. Ns0's largest effect is observed for low values of It*Ns and becomes basically unnoticeable for large values, indicating the value of Ns0 is of little consequence, provided a sufficient number of It*Ns models have been considered.

In summary, It*Ns, Nr, Ns0 control the behavior of the inversion algorithm. Of these parameters, It*Ns models is the most significant, followed by Nr and then Ns0. While not strictly a tuning parameter, the number of trials performed (Ntrials = 10 in this study) is shown to have a significant effect on the inversion's convergence, even when a large number of models are considered. It is recommended that Ntrials be at least three. However, since this parameter is tied to the non-linearity of the inverse problem being considered it is expected to vary on a case-by-case basis. Meaning if significant variability is observed after performing three trials additional trials may be necessary to ensure the selection of the best possible solution. An Nr of $\approx$100 is recommended, as it is believed to be a reasonable compromise between faster convergence and an increase in the production of outliers. If outliers are seen to persist when using Nr $\approx$100, Nr may be increased to make the inversion more explorative, however It*Ns will then also need to be increased. Against intuition, the utilization of a large number of models (\textgreater30,000 for Nr=100) will generally not produce a lower misfit than considering a fewer but sufficient number of models. However, as this test was setup under ideal conditions it represents a lower bound and therefore a value of It*Ns of 50,000 in conjunction with Nr=100 is recommended. The generality of the recommendation of using Nr=100 and It*Ns=50,000 is confirmed through the validation of the workflow discussed next. As Ns0 is shown to have a small effect, any Ns0 greater than Nr (to ensure that the NA has a sufficient number of starting models to begin) is recommended. The values recommended here have been set as the defaults in the DesignSafe-CI application for performing batch-style surface wave inversions, \emph{SWbatch}, \citep{vantassel_jpvantasselswbatch_2020} for the user's convenience.

\section{Validation of Inversion Workflow}

To validate the proposed workflow and make recommendations about the parameterization approaches previously discussed, 12 sets of synthetic experimental dispersion data were developed, inverted, and the resulting Vs profiles compared.

\subsection{Selection of Inversion Targets}
\begin{figure}[!ht]
	\includegraphics[width=\textwidth]{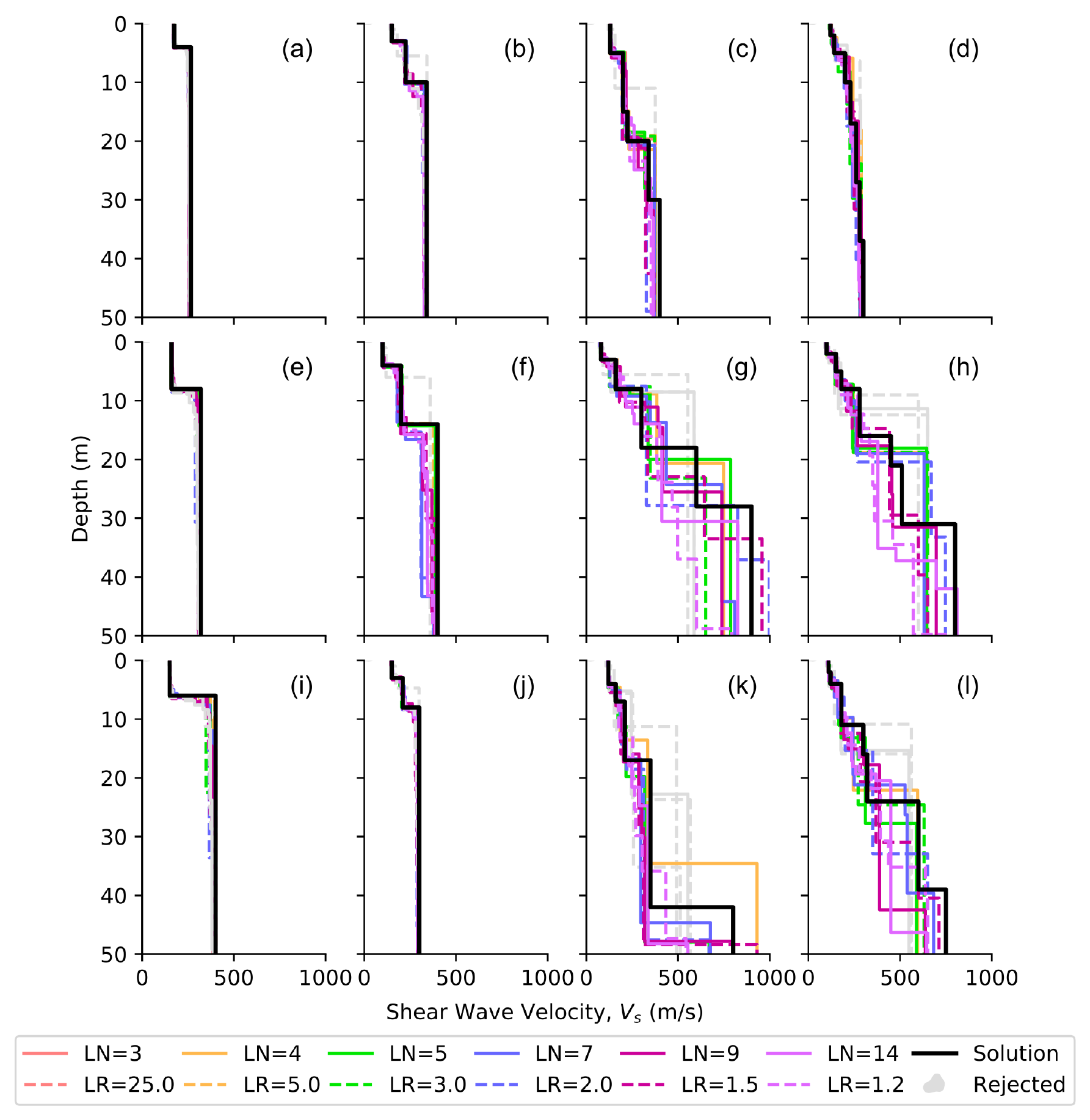}
	\caption{Comparison of the lowest dispersion misfit shear-wave velocity (Vs) profile from each parameterization with the true solution. The Layering by Number (LN) and Layering Ratio (LR) parameterizations which contain the same number of layers share the same color and are adjacent to one another in the figure's legend (e.g., LN=5 and LR=3.0 are both 5-layered parameterizations). The LN and LR parameterizations are distinguishable by their solid and broken lines, respectively. Rejected profiles are shown in grey.}
	\label{fig:setvs}
\end{figure}

The experimental dispersion data were generated from 12 synthetic velocity models. The 12 velocity models were designed to belong to four categories, with each category shown as a column of Figure \ref{fig:setvs}.  The first column [panels (a), (e), and (i)] represents the 2-layer category, the second column [panels (b), (f), and (j)] the 3-layer category, the third column [panels (c), (g), and (k)] the 5-layer category, and the fourth column [panels (d), (h), and (l)] the 7-layer category. Three velocity models of increasing complexity were designed for each of the four layering categories. The first model in each layering category [panels (a)-(d)]  was selected to represent a profile with a gradual increase in Vs following approximate relationships between Vs and the mean-effective stress \citep{menq_dynamic_2003, lin_variability_2008}. The second model in each category [panels (e)-(h)] was designed to include one or more large impedance contrasts where the large contrasts were selected to be realistic representations of transitions from soft to dense soil, dense soil to weathered rock, and weathered to competent rock.  The third model in each category [panels (i)-(l)] was designed to include a mixture of these two approaches. The compression wave velocity (Vp) for these models was selected to be consistent with unsaturated soils and rock (i.e., Poisson's ratio between 0.2 and 0.35). The mass density for these models was kept fixed at 2000 kg/m\textsuperscript{3}. Synthetic fundamental-mode Rayleigh wave mean experimental dispersion data for each model was calculated via the Thomson-Haskell forward problem. The synthetic experimental dispersion data was assumed to be normally distributed in terms of Rayleigh-phase velocity with a coefficient of variation of 0.05 for reasons discussed previously. The data were resampled in terms of log-wavelength between 2 and 150 m using 20 samples. 

\subsection{Selection of Inversion Parameterizations and Tuning Parameters}
Each synthetic model was parameterized as follows for inversion. Vp was limited between 130 and 2000 m/s with layering equal to that of Vs. To reduce complexity, the thickness of each Vp layer was linked to that of Vs. Vs was limited between 70 and 1000 m/s and was parameterized using LN's = 3, 4, 5, 7, 9, and 14 as well as LR's = 25.0, 5.0, 3.0, 2.0, 1.5, and 1.2. Note that the LRs were selected so that they would have the same number of layers (although different potential layer boundaries) as their LN counterparts to allow for a comparison of the two parameterizations. For example, LR = 25, which is an atypical LR and outside of the recommended range (see recommended LRs above), was included because it yielded a parameterization with three layers and provided a useful basis of comparison with LN = 3, which is within the recommended range (see recommended LNs above). Mass density was held fixed at 2000 kg/m\textsuperscript{3}. Poisson's ratio for all layers was allowed to vary between 0.2 and 0.5 and was not constrained to decrease with depth. Each parameterization was inverted using Ntrials = 10, Ns0 = 10,000, Nr = 100, It = 250, and Ns = 200 (i.e., 60,000 models per trial; 0.6 million per parameterization). Meaning, across the 12 parameterizations considered for each synthetic model, 7.2 million models were explored. Thus, across the 12 synthetic models considered 86.4 million trial inversion models were searched to develop these results. Note, using Ns0 = 10,000 was done, even though Ns0 has only a small effect on convergence (refer to Figure \ref{fig:tunestudy}), due to the negligible computation cost of these models. Dinver is capable of solving Ns0 models approximately 10 to 100 times faster than those searched with the NA (i.e., It*Ns). Using a large value for Ns0 helps ensure that a good set of models have already been evaluated before the first NA iteration and minimizes the potential for becoming stuck in a local minima at minimal computational cost.

\subsection{Detailed explanation of a single example}

\begin{figure}[!ht]
	\includegraphics[width=\textwidth]{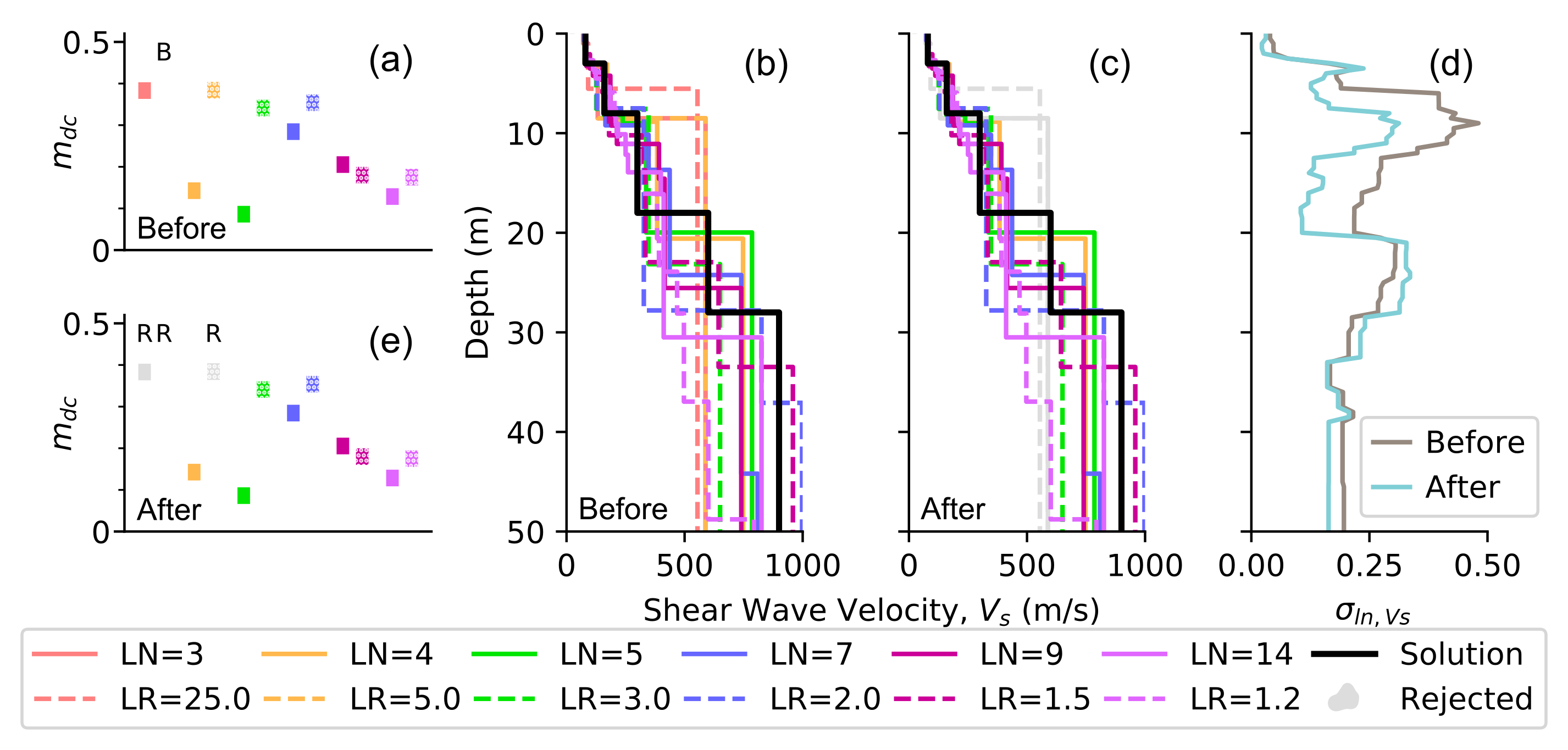}
	\caption{Presentation of a single detailed example. The minimum dispersion misfit ($m_{dc}$) achieved for each parameterization before and after rejection are shown in panels (a) and (e), respectively. The $m_{dc}$ values not shown in panel (a) were beyond (B) the vertical range and are designed with a B. The $m_{dc}$ values in panel (e) that were rejected (R) are shown in grey and designated with an R. Layering by Number (LN) and Layering Ratio (LR) parameterizations with the same number of layers share the same color (e.g., LN=5 and LR=3.0 are both 5-layered parameterizations), and can be distinguished by their solid and hatched texture, respectively. The corresponding lowest misfit Vs profiles before and after rejection are shown in panels (b) and (c), respectively. The LN and LR parameterizations containing the same number of layers share the same color and are distinguishable by their solid and broken line style, respectively. Rejected profiles are shown in light grey. Panel (d) compares the $\sigma_{ln,Vs}$ for the profiles before and after rejection.}
	\label{fig:singlevs}
\end{figure}

To facilitate a meaningful discussion about the process of developing realistic Vs profiles, we first examine a single example from the 12 synthetic velocity models shown in Figure \ref{fig:setvs}. Following a detailed explanation of this single example, we will return to Figure \ref{fig:setvs} to discuss all of the results. The synthetic model chosen for initial detailed discussion is a five layer model with strong impedance contrasts (refer to Figure \ref{fig:setvs}g). This synthetic model and its corresponding inversion results are shown in Figure \ref{fig:singlevs}. Note that at present only the single best (i.e., lowest misfit) result is shown for each parameterization (i.e., the lowest misfit Vs profile obtained from the 0.6 million trial models). It can be observed that not all parameterizations perform equally. For example, when examining the $m_{dc}$  values (refer to Figure \ref{fig:singlevs}a) the LR=25.0 layering parameterization is shown to under-perform its counterparts, which generally range from 0.1 to 0.4. In fact, the LR=25 misfit is beyond (B) the vertical scale and therefore shown with a ``B". When examining the entire suite of Vs profiles (refer to Figure \ref{fig:singlevs}b) for trends, LN=3, LR=25, and LR=5.0 are seen to not follow the same trends as the other Vs profiles, tending to yield greater Vs than the other parameterizations between 5 m and 20 m. This behavior indicates that these parameterizations were under-parameterized (i.e., contained too few layers and/or too restrictive layer boundaries) and could not achieve good fits to the true solution. Note that while the $m_{dc}$  for LR=25 is relatively high compared to the other layering parameterizations, the $m_{dc}$  for LN=3 and LR=5.0 are not significantly higher than the others. Thus, by examining the $m_{dc}$  values alone one could not know that the resulting Vs profile did not well fit the true model. As such, the $m_{dc}$  values and relative agreement/disagreement of the resulting Vs profiles for different parameterizations must be used together to identify Vs profiles that may potentially not be representative of the true subsurface. For these two reasons (i.e., relatively high $m_{dc}$  as compared to the other profiles coupled with inconsistent Vs profiles compared to the others) the Vs profiles resulting from LN=3, LR=25, and LR=5.0 were rejected (R). The suite of profiles after rejection are shown in Figure \ref{fig:singlevs}c, with the rejected profiles shown in light grey and indicated with an ``R" in the summary of $m_{dc}$  in Figure \ref{fig:singlevs}e. It is important to note that these profiles were not rejected based on their relationship to the solution (as this is never known in practice), but only in regards to their $m_{dc}$  values and Vs profiles when viewed in comparison to the other parameterizations. Figure \ref{fig:singlevs}d shows the lognormal standard deviation of Vs ($\sigma_{ln,Vs}$) from the profiles before and after rejection. It is clear that if the rejected profiles had been considered as viable solutions representative of the subsurface they would have significantly biased the resulting estimates of uncertainty, as their removal reduces $\sigma_{ln,Vs}$ from $\approx$0.4 to $\approx$0.25 between 5 m and 20 m. To generalize, we propose that the results from a trial parameterization can be rejected if: (1) it has a relatively high misfit compared to the other parameterizations considered, and (2) its Vs profiles are true outliers relative to the others, particularly when the parameterization resides at the edge of the range of parameterization complexity considered (i.e., lowest/highest numbers of trial layers). All other results, while variable, must be considered as possible representations of the subsurface. 

With those underperforming parameterization removed it is now easier to comment on the behavior of the accepted profiles (refer to Figure \ref{fig:singlevs}c). First, the agreement between the inverted profiles and the true solution tends to decrease with depth. This, as mentioned previously, should be expected due to the mixed-determined nature of the inverse problem. Second, while no profile exactly matches the true solution, by considering multiple parameterizations the resulting profiles are shown to generally bound the true solution. This further emphasizes the importance of considering multiple parameterizations when performing surface wave inversion. Selecting only a single parameterization, even if it happens to have the correct number of layers and a low $m_{dc}$  (e.g., LN=5), may result in a misleading interpretation of the subsurface and a failure to communicate the uncertainties present in the analysis.

\subsection{Evaluation of Dispersion Misfit}

\begin{figure}[!ht]
	\includegraphics[width=\textwidth]{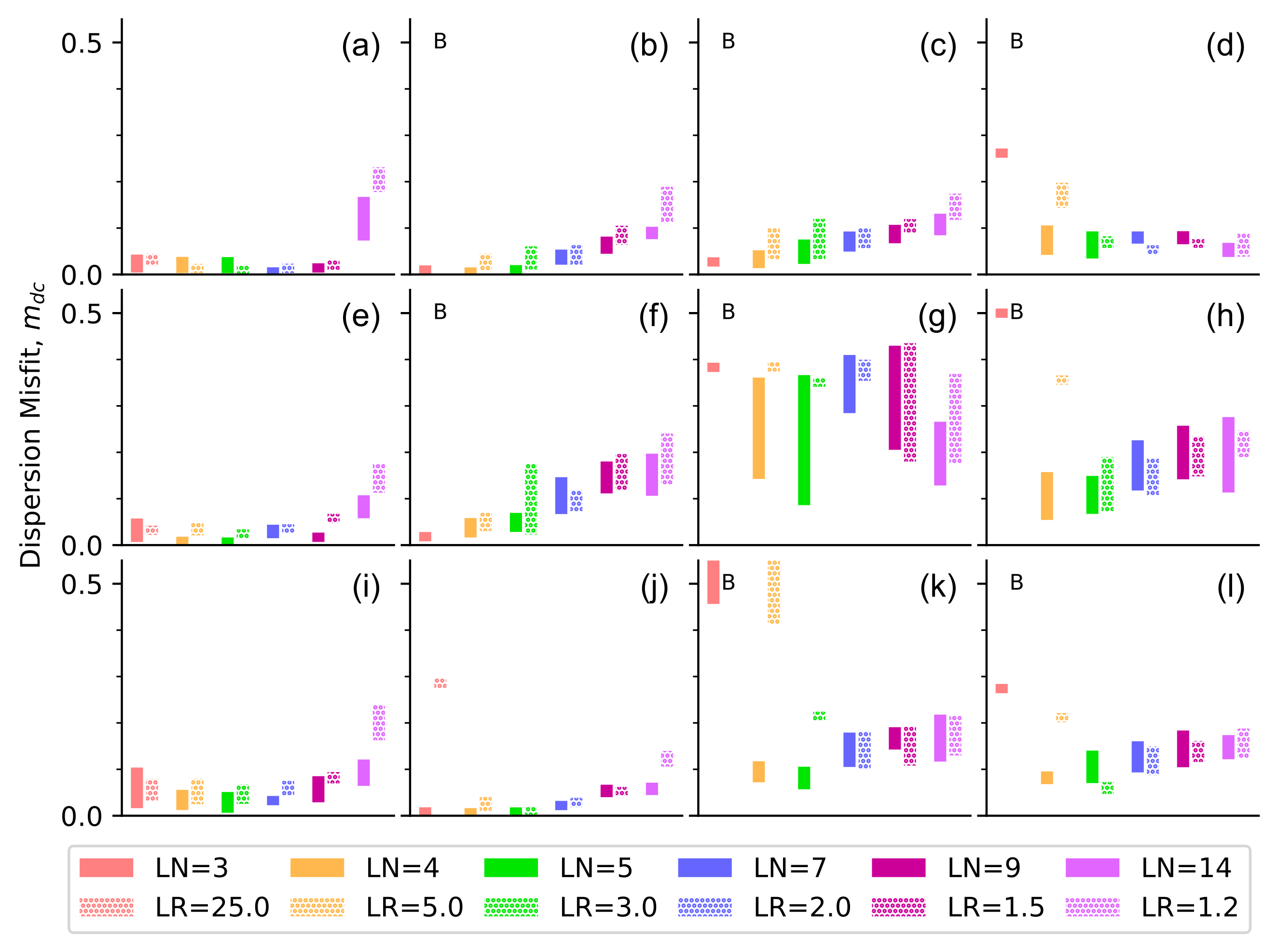}
	\caption{Dispersion misfit ($m_{dc}$) values achieved during inversion of 12 different synthetic models, (a) -- (l). The vertical bars denote the range of the minimum $m_{dc}$ for each inversion parameterization (one per trial, ten trials per parameterization). The Layering by Number (LN) and Layering Ratio (LR) parameterizations containing the same number of layers share the same color and are adjacent to one another in the figure and legend (e.g., LN=5 and LR=3.0 are both 5-layered parameterizations). LN and LR are distinguishable by their solid and hatched texture, respectively. Parameterizations with misfit ranges beyond (B) the vertical scale of 0.5 are not shown and instead denoted with a B.}
	\label{fig:mdc}
\end{figure}

Now, we discuss in detail the results from inverting all 12 synthetic Vs models, beginning with the $m_{dc}$  trends. The ranges of the minimum dispersion misfits obtained for each parameterization are summarized in Figure \ref{fig:mdc}. Recall, each model was inverted using 12 different parameterizations (6 LN-type and 6 LR-type), where each parameterization was inverted independently 10 times (i.e., Ntrials = 10) using 60,000 models per trial. The range of misfit values shown in Figure \ref{fig:mdc} are the single lowest misfit value from each of the 10 trial inversions. The columns and rows correspond to those models previously discussed in regards to Figure \ref{fig:setvs}. The $m_{dc}$  shown along the y-axis of each panel is the normalized root-mean-square error proposed by \citet{wathelet_surface-wave_2004}. A nice feature of this misfit function is its physical interpretation; that the misfit represents (on average, across frequency or wavelength) the distance in numbers of standard deviations away from the mean that the theoretical dispersion curve resides relative to the experimental data. Thus, a misfit value less than 1.0 essentially means that the theoretical dispersion curve falls within the $\pm$1 standard deviation bounds of the experimental data. As noted by \citet{cox_layering_2016}, misfit values greater than 1.0 are generally considered poor or unacceptable. However in practice, the complexity of the experimental data (mode interpretation issues, frequency/wavelength gaps, large vs. small dispersion uncertainty, etc.) plays a significant role in the misfit value that can be achieved at any given site. As such, misfit values from different sites generally cannot be compared directly with one another for a measure of the overall inversion quality from site-to-site. Rather, the misfit values can simply be used to guide relative judgements about the quality of certain layered earth models relative to others at the same site \citep{griffiths_surface-wave_2016}.

In order to visualize the massive amounts of data generated in this study, we will primarily focus first on finding the minimum misfit model for each parameterization (i.e., those models that best explain the mean trend of the experimental dispersion data), although in practice (as will be discussed briefly later) efforts should be made to find all/many models which explain both the mean trend and the uncertainty in the experimental dispersion data. From Figure \ref{fig:mdc}, we observe several general trends. First, that the range of minimum misfit values attained across the ten trials for any given parameterization (i.e., the height of each vertical bar) can vary significantly, indicating that it is important to consider multiple trials for each parameterization when performing surface wave inversions and that the effect of doing so is highly problem dependent. Second, the vast majority of misfit values are below 0.5 with many of those below 0.1; meaning that any of these results would generally be found to be acceptable in practice. Third, in terms of the minimum misfit attained, the LN parameterizations tend to out-perform their LR counterparts, however, there is no clear advantage of one parameterization over another for use in all cases. Fourth, those parameterizations that have numbers of layers consistent with the true model tend to perform better (i.e., produce lower misfits) than those parameterizations with numbers of layers inconsistent with the true model. Fifth, the more complex models tend to have higher and more variable misfits than the more simplified models; further emphasizing the value of considering multiple parameterizations and multiple trials per parameterization on realistic models. And finally, that all parameterizations do not produce models of equal quality (i.e., minimum misfit). These factors combined indicate that in order to find the best solution(s) and investigate uncertainty multiple parameterizations must be considered. To do this properly, there must be a systematic approach to reject under-performing parameterizations, such as that proposed in the previous section, so as to not bias the results with unrealistic profiles which do not well-represent the subsurface.

\subsection{Qualitative Comparison of Dispersion}

\begin{figure}[!ht]
	\includegraphics[width=\textwidth]{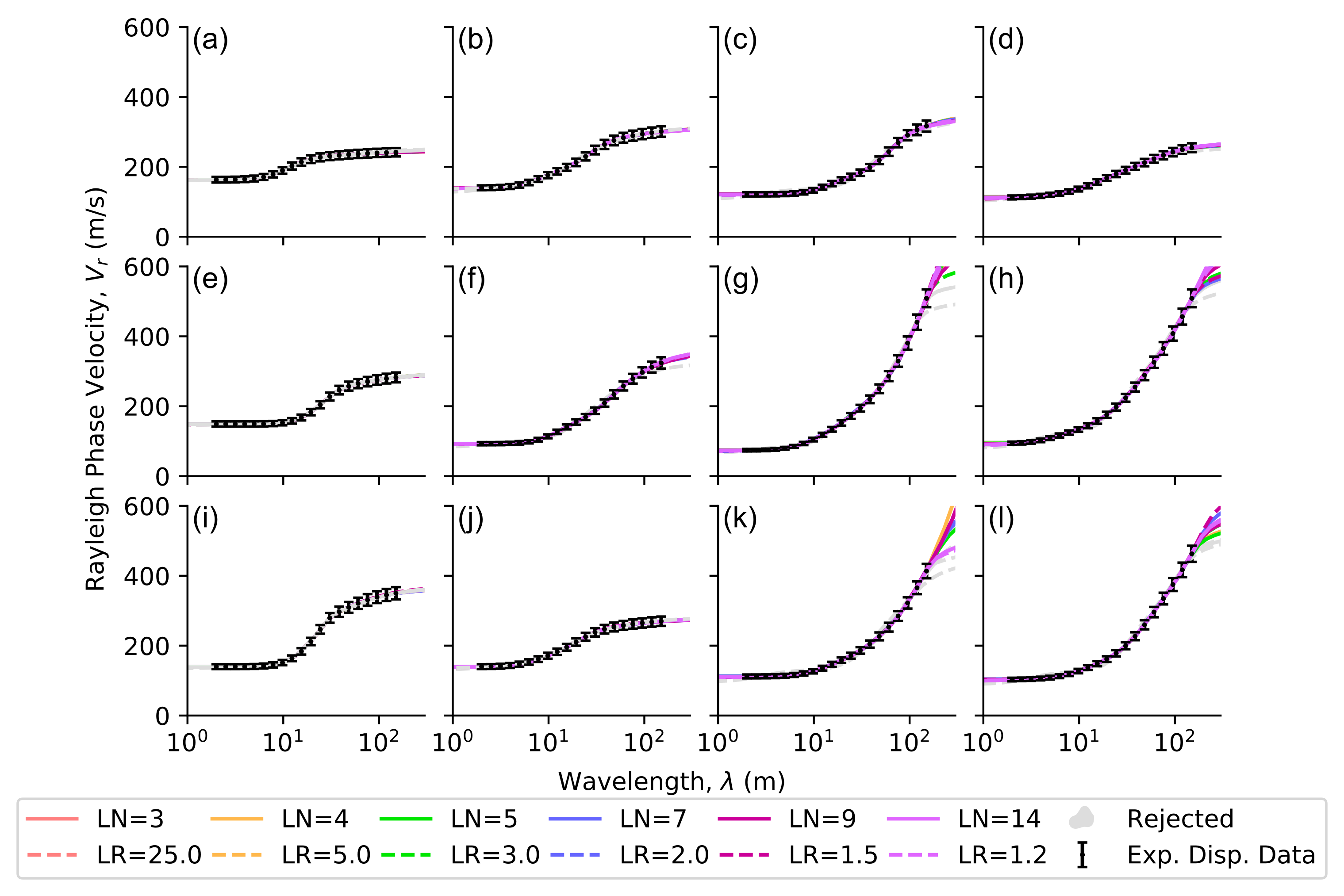}
	\caption{Comparison of the lowest misfit ground model's fundamental mode Rayleigh wave (R0) theoretical dispersion curve to the experimental dispersion data for each Vs parameterization for the 12 different synthetic models, (a) -- (l). The Layering by Number (LN) and Layering Ratio (LR) parameterizations which contain the same number of layers share the same color and are adjacent to one another in the figure's legend (e.g., LN=5 and LR=3.0 are both 5-layered parameterizations). The LN and LR parameterizations are distinguishable by their solid and broken lines, respectively. Results from rejected (R) parameterization are shown in light grey.}
	\label{fig:setdc}
\end{figure}

In Figure \ref{fig:setdc}, the lowest $m_{dc}$  theoretical dispersion curve for each parameterization is shown in comparison to the experimental dispersion data. The dispersion curves shown here correspond to the $m_{dc}$  at the bottom of each vertical bars shown in Figure \ref{fig:mdc}. Stated another way, these dispersion curves are the lowest misfit curves produced across all 10 trial inversions for each parameterization considered. The columns and rows correspond to those shown in Figure \ref{fig:setvs} and 7. In general, the theoretical dispersion curves are shown to visually fit the experimental dispersion data well, with only a few exceptions. These exceptions are most evident at long wavelengths where the theoretical curve diverges from the experimental data [e.g., panels (g) and (k)]. From Figure \ref{fig:setdc}, we also observed the general shapes of the experimental dispersion data, which corresponds to two categories. The first category is distinguished by ``S"-shaped experimental dispersion data, which includes panels (a)-(d), (e)-(f), and (i)-(j). Recall that these models correspond to gradual increasing Vs profiles [panels (a)-(d)], 2-layered Vs profiles [panels (a), (e), and (i)], and 3-layered Vs profiles [panels (b), (f), and (j)]. The second category is distinguished by ``L"-shaped experimental dispersion data, which includes panels (g)-(h) and (k)-(l). Recall that these models corresponded to Vs profiles with large impedance contrasts and a greater numbers of layers (i.e., \textgreater4). For these models the experimental dispersion data at long wavelengths does not flatten towards the half-space Rayleigh wave velocity, as it does with those models in category 1 (i.e., ``S"-shaped). As such, the maximum wavelength (i.e., 150 m) is not sufficiently long to capture the half-space velocity. The effect of the category 2 dispersion data not ``flattening-out" at long wavelengths will be discussed in context to the goodness of the resulting Vs profiles derived from inversion.

\subsection{Qualitative Comparison of Vs}

Figure \ref{fig:setvs} compares the lowest dispersion misfit Vs profiles from each parameterization with the true solution. Several observations can be made regarding Figure \ref{fig:setvs}. First, the Vs profiles rejected based on the two criteria previously discussed in regards to the detailed example shown in Figure \ref{fig:singlevs} do not satisfactorily match the true Vs model and therefore were rightly discarded. Second, inversion-derived Vs profiles in category 1 (``S"-shaped dispersion data) illustrate excellent fits to the true model, after under-performing parameterizations have been rejected based on the two proposed criteria (i.e., $m_{dc}$  and Vs trend). Third, the Vs profiles in category 2 (``L"-shaped dispersion data) do not fit the solution as well as those profiles in category 1. The inability of these profiles to exactly capture the true profile is the result of at least two contributing factors: (1) these models contain the largest numbers of layers considered and are thereby most affected by the inverse problem's non-uniqueness, and (2) these model's experimental dispersion data does not extend to long enough wavelengths to capture the half-space Rayleigh wave velocity, thereby making accurate predictions of the last layer's stiffness and depth difficult (if not impossible in some cases). Yet, despite these complicating factors we observe that these profiles are still able to capture the trend of increasing velocity with depth in the upper 30-40 m. Additionally, by using multiple parameterizations the relative uncertainty of the ``L"-shaped category profiles as compared to the ``S"-shaped category is immediately apparent.

\subsection{Quantitative Comparison of Vs}

\begin{figure}[!t]
	\includegraphics[width=\textwidth]{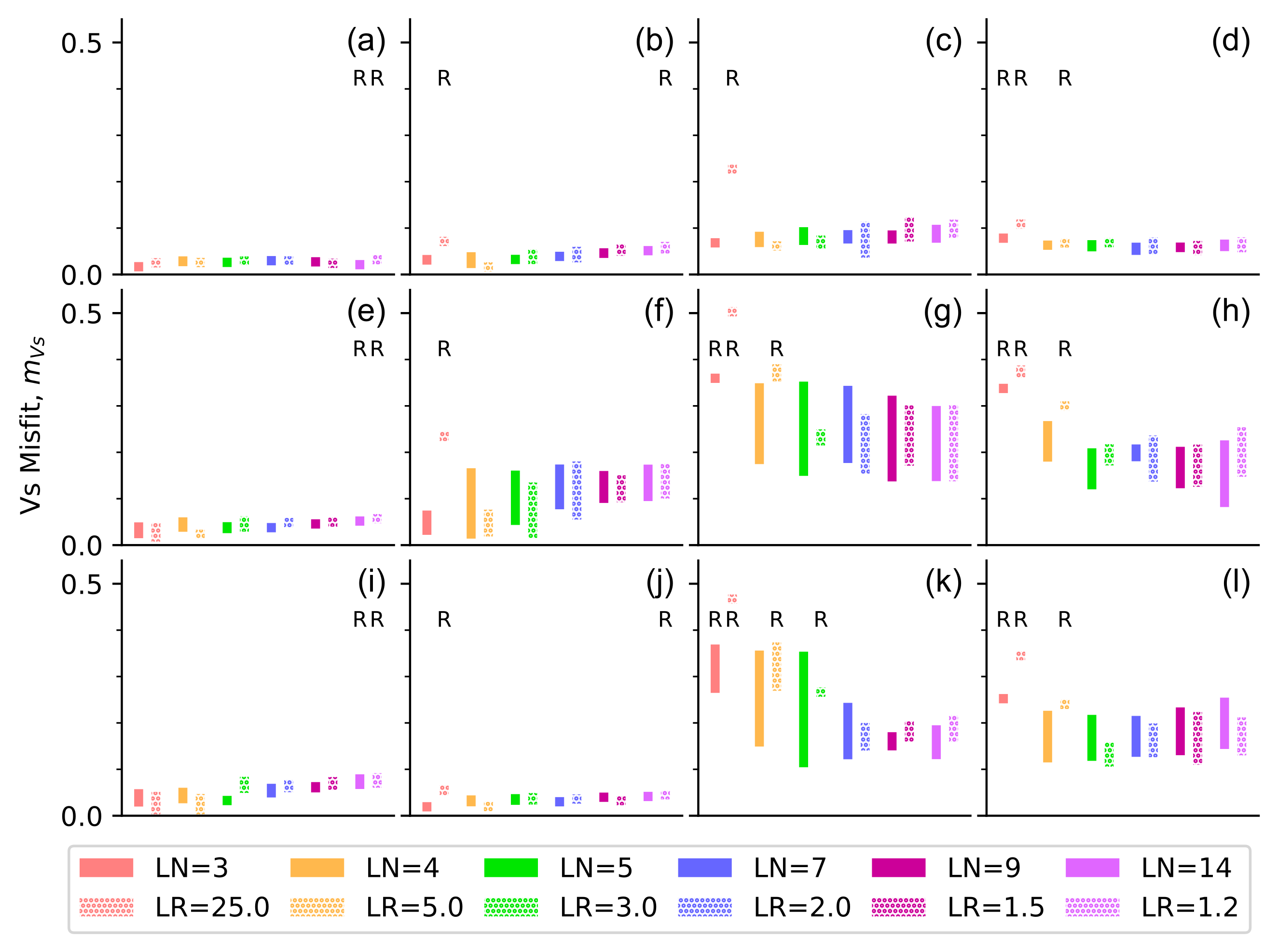}
	\caption{Shear wave velocity (Vs) misfit ($m_{Vs}$) values achieved during inversion of 12 different synthetic models, (a) -- (l). The vertical bars denote the range of $m_{Vs}$ values (one per trial, ten trials per parameterization) for the best model from each trial inversion for each inversion parameterization. The Layering by Number (LN) and Layering Ratio (LR) parameterizations containing the same number of layers share the same color and are adjacent to one another in the figure and legend (e.g., LN=5 and LR=3.0 are both 5-layered parameterizations). LN and LR are distinguishable by their solid and hatched texture, respectively. Parameterizations that were rejected (R) using the two proposed criteria are shown with an ``R" above/below their respective vertical bars.}
	\label{fig:mvs}
\end{figure}

To quantify the goodness of fit of each profile to the true solution a Vs misfit ($m_{Vs}$) following the form of an average percent difference is proposed in Equation 1:

\begin{equation}
m_{Vs} = \frac{1}{N} \sum_{i=1}^{N} \frac{|Vs_{i,inversion} - Vs_{i,solution}|}{Vs_{i,solution}}
\end{equation}

where N is the desired number of depth discretizations for the Vs profiles, $Vs_{i,inversion}$ is the value of Vs for the inversion-derived profile at the ith depth discretization, and $Vs_{i,solution}$ is the value of Vs for the solution at the ith depth discretization. For this study, the profiles were discretized into 0.5 m increments from 0 to 50 m. The result of the application of this Vs misfit function is shown in Figure \ref{fig:mvs}. The columns and rows correspond to those used previously in Figures 5, 7, and 8. Each panel shows the range of $m_{Vs}$  across the lowest dispersion misfit Vs profiles obtained from ten trials performed for each parameterization. Again, a clear distinction between those profiles in category 1 (``S"-shaped dispersion data; models a, b, c, d, e, f, i, and j) and category 2 (``L"-shaped dispersion data; models g, h, k, and l) can be observed. The best Vs profiles in category 1 exhibit $m_{Vs}$  below 0.1, whereas the best Vs profiles for category 2 exhibit $m_{Vs}$  generally greater than 0.1, indicating quantitatively what was observed qualitatively in Figure \ref{fig:setvs}. To examine how the criteria for rejecting the Vs profiles compares quantitatively to the goodness of fit, the rejected parameterizations have been denoted with an ``R". We observe that the profiles rejected based on the two proposed criteria have corresponding large $m_{Vs}$  which, if not rejected, would bias the results with unrealistic profiles. Hence, this is quantitative evidence that these profiles were rightly rejected. Additionally, we propose this Vs misfit function in hopes that researchers who wish to validate their inversion algorithms in future studies may use this set of 12 synthetic models, which have been archived on the DesignSafe-CI \citep{vantassel_surface_2020}, and this misfit function to directly compare their results with those shown here.

\section{Accounting for Uncertainty}

\begin{figure}
	\includegraphics[width=\textwidth]{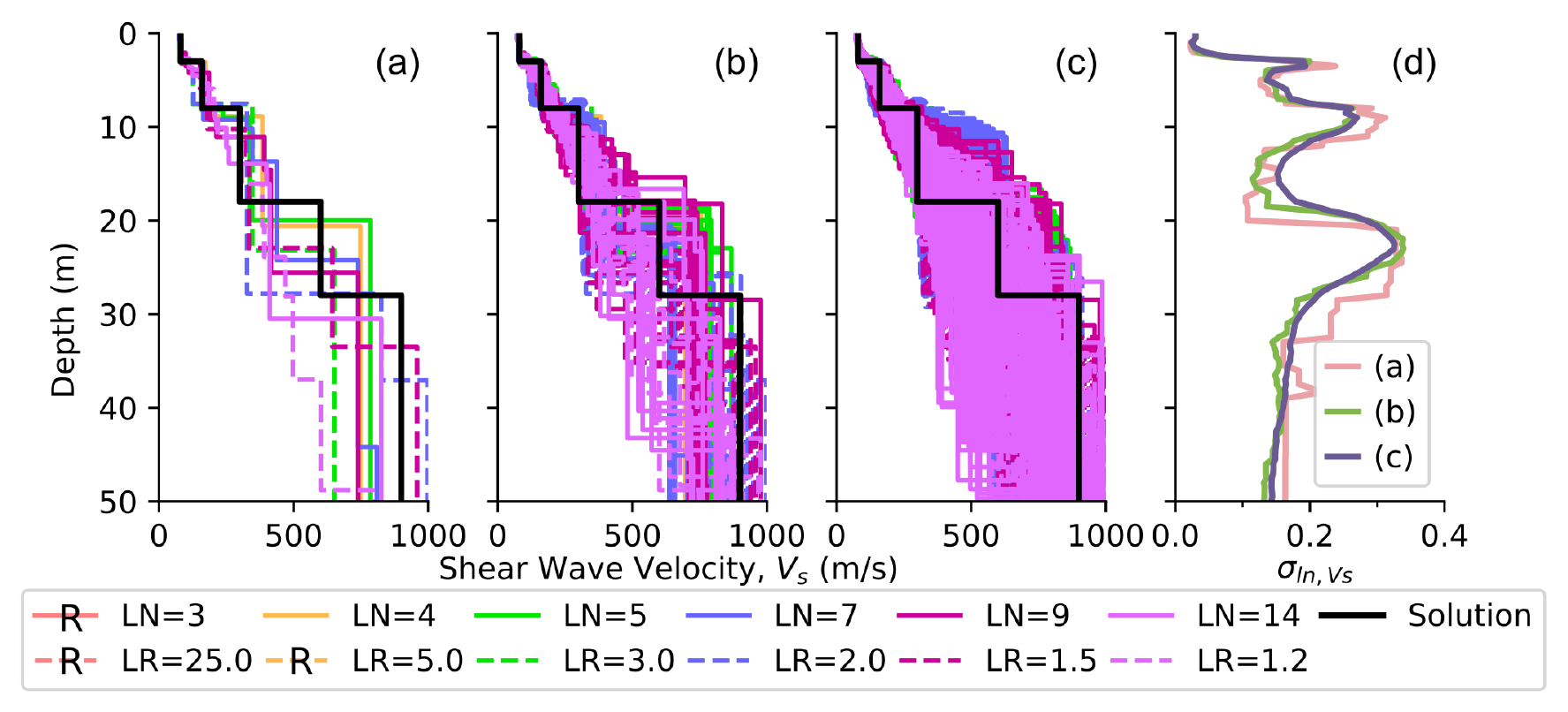}
	\caption{Approximate methods for accounting for Vs uncertainty applied to dataset g (i.e., a high-variance dataset). The methods include: (a) considering the single best model from each acceptable inversion parameterization, (b) considering the 10 lowest misfit models (across all trials) from each acceptable inversion parameterization, and (c) considering the 100 lowest misfit models (across all trials) from each acceptable inversion parameterization. Panel (d) shows the $\sigma_{ln,Vs}$ corresponding to each panel previously discussed. The parameterizations which have been rejected (R) are indicated with an R in the figure legend.}
	\label{fig:highvar}
\end{figure}

Due to the non-uniqueness of the inverse problem, it is generally unreasonable to expect that surface wave inversion can produce a single unique result that exactly represents the true profile. This is shown qualitatively in Figure \ref{fig:setvs} for models with realistic complexity in terms of numbers of layers and strong stiffness contrasts (models g, h, k, and l). Still, even in these challenging cases all accepted profiles in Figure \ref{fig:setvs} are shown to: (1) reliably capture the trends of increasing velocity with depth and (2) bound the true solution when multiple parameterizations are considered. With these considerations in mind, it is of paramount importance to try to quantify the uncertainty in the inversion-derived Vs profiles. There have been several ways suggested in the literature to quantify Vs uncertainty resulting from surface wave inversions. These methods generally involve selecting suites of inversion models, from which statistics are calculated. These suites can be selected in various ways, but most commonly include some ``reasonable" number of the best/lowest misfit models (e.g., the best 100 or 1000 models). Regardless, no single method has been accepted into practice, and therefore the selection of models representative of the inversion's uncertainty is left to the analyst. What we propose here is an approximate method for quantifying Vs uncertainty that is simple and can be performed immediately following the inversion after the rejection of low-quality parameterizations. The application of the method to a high-variance dataset from category 2 (i.e., dataset g) is illustrated in Figure \ref{fig:highvar}. Figure \ref{fig:highvar}a shows the minimum misfit profiles from each parameterization after rejection. Rejected parameterizations are indicated with an R in the figure legend. The variability of the nine accepted lowest misfit profiles represents the inter-parameterization uncertainty, which is quantified in Figure \ref{fig:highvar}d using the lognormal standard deviation of Vs ($\sigma_{ln,Vs}$ ). To visualize the effects of incorporating some of the intra-parameterization uncertainty, the 10-lowest misfit profiles from each of the accepted parameterizations are shown in Figure \ref{fig:highvar}b, and their $\sigma_{ln,Vs}$  in Figure \ref{fig:highvar}d. Note that the value of $\sigma_{ln,Vs}$  does not change significantly, indicating that for this case the inter-parameterization uncertainty is more significant than the intra-parameterization uncertainty, again highlighting the importance of considering multiple parameterizations. To investigate further, the 100-lowest misfit profiles from each of the accepted parameterizations are shown in Figure \ref{fig:highvar}c, and their $\sigma_{ln,Vs}$  in Figure \ref{fig:highvar}d. Figures 10a, b, and c give a qualitative sense as to which parts of the profile are the most uncertain (half-space velocity and layer boundaries) and which are well constrained (near-surface velocity and increase of velocity with depth). These observations are mirrored quantitatively in Figure \ref{fig:highvar}d, where $\sigma_{ln,Vs}$  is smallest near the surface and increases with depth. The bulges in $\sigma_{ln,Vs}$  are associated with the locations of layer boundaries and quantitatively show that these locations are less certain. This is not to say that layer boundaries are highly uncertain as the increased values of $\sigma_{ln,Vs}$  may imply, but rather the calculation of $\sigma_{ln,Vs}$  as typically performed mixes the uncertainty of the layer boundaries with that of Vs. Thereby making it appear as if these regions of the model are highly uncertain, when in reality it is a side effect of the uncertainty of the layer boundary and the procedure for calculating  $\sigma_{ln,Vs}$.

\begin{figure}
	\includegraphics[width=\textwidth]{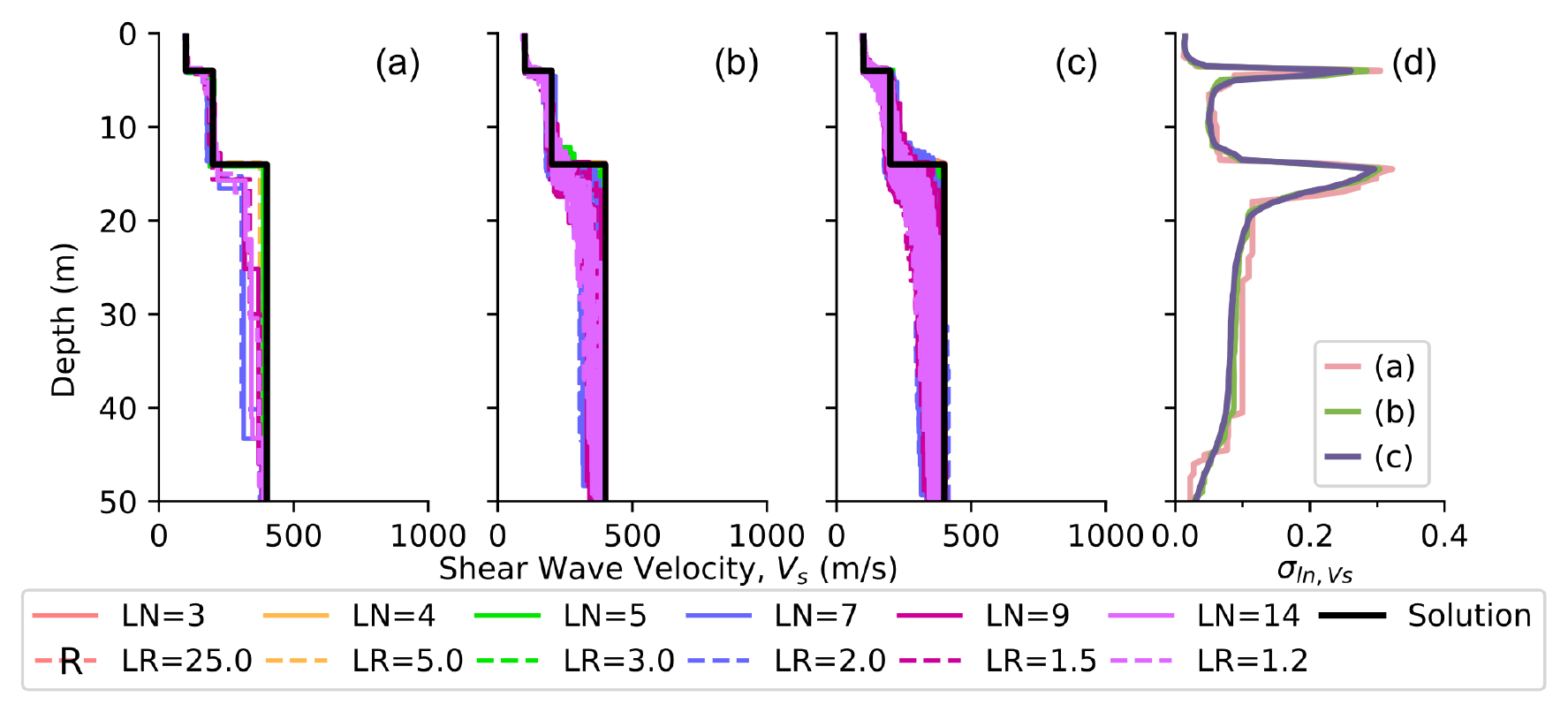}
	\caption{Approximate methods for accounting for Vs uncertainty applied to dataset f (i.e., a low-variance dataset). The methods include: (a) considering the single best model from each acceptable inversion parameterization, (b) considering the 10 lowest misfit models (across all trials) from each acceptable inversion parameterization, and (c) considering the 100 lowest misfit models (across all trials) from each acceptable inversion parameterization. Panel (d) shows the $\sigma_{ln,Vs}$ corresponding to each of panels previously discussed. The parameterizations which have been rejected (R) are indicated with an R in the figure legend.}
	\label{fig:lowvar}
\end{figure}

To contrast the high-variance dataset in Figure \ref{fig:highvar}, Figure \ref{fig:lowvar} presents similar plots for a low-variance dataset from category 1 (i.e., dataset f). Figure \ref{fig:lowvar} shows significantly less variance in the Vs profiles than was observed in Figure \ref{fig:highvar}, which implicitly expresses that these profiles are more certain than those in Figure \ref{fig:highvar}. The $\sigma_{ln,Vs}$  in Figure \ref{fig:lowvar}d echoes this qualitative assessment quantitatively, with the average uncertainty in Vs being reduced from $\approx$0.2 for the high-variance dataset to $\approx$0.1 in the low-variance dataset. Again, the inter-parameterization uncertainty is shown to be significantly greater than the intra-parameterization uncertainty, with large spikes in the uncertainty being related to the location of layer boundaries. To provide context to the values of $\sigma_{ln,Vs}$  presented here, typical values for site specific applications presented in the literature include 0.27 -- 0.37 \citep{toro_probabilistic_1995}, 0.5 at sites with little to no information to 0.25 (and potentially less) for very thoroughly characterized sites \citep{epri_seismic_2012}, and 0.15 in the upper 50 m and 0.22 at depth for low variability sites \citep{stewart_guidelines_2014}. These literature values illustrate that even for the high-variance dataset presented here (refer to Figure \ref{fig:highvar}) though the results may appear highly uncertain the corresponding values of $\sigma_{ln,Vs}$  are equal to or less than those currently being assumed in practice for well-characterized sites.

Based on the examples presented, if a few Vs profiles are desired for use in future engineering analyses (e.g., site response analyses) we recommend using each of the minimum misfit profiles from all acceptable inversion parameterizations as an approximate approach to incorporating Vs uncertainty. To communicate the inter- and intra-parameterization uncertainty, we recommend showing the 100 lowest misfit profiles for each acceptable inversion parameterization as a qualitative measure along with $\sigma_{ln,Vs}$  as a quantitative measure. The ability to easily plot the 100 (or 1,000 or 10,000) lowest misfit profiles and their associated $\sigma_{ln,Vs}$  has been included as a part of \emph{SWprepost}.

\section{Recommendations and Conclusions}
The recommendations and conclusions of this study are summarized in the following proscriptions:
\begin{itemize}
	\item When resampling experimental dispersion data log-wavelength resampling should be preferred to log-frequency sampling, and linear-frequency resampling should not be used.
	\item When developing trial inversion parameterizations all high-quality site-specific information (geology, boring logs, etc.) should be used to constrain the layering and develop reasonable parameter limits. However, the user should be cautious not to overly constrain the inversion's parameter space and thereby bias the inversion results.
	\item When parameterizing Vs, if limited site specific information is available or site specific information does not extend to a sufficient depth, the layering by number (LN) and layering ratio (LR) parameterizations are recommended. Note, while it is not addressed specifically in this work, the authors' experience applying this methodology to real datasets indicates that LN tends to perform better when fewer layers are expected in the subsurface whereas LR tends to perform better when many layers are expected. Therefore, we recommend a hybrid approach which includes low LN's such as 3, 4, 5, and 7 be combined with low LR's (i.e., large numbers of layers) such as 3.0, 2.0, 1.5, and 1.2 for accounting for parameterization uncertainty.
	\item When parameterizing in terms of Vp, we recommend using the same number of layers for Vp as Vs and linking the thickness of Vp to Vs to reduce parameterization complexity. We strongly advise against assuming Vp or using a constant Poisson's ratio without site specific justification, as doing so may strongly bias the resulting Vs profiles, especially for materials with high Vp (i.e., saturated soils, inter-mediate geomaterials, and rock). Furthermore, if the depth to saturated soils, gravel, or rock is known (or can be reasonably estimated) these ranges in terms of depth should be preferred over allowing Vp to range widely, as was necessary in this study because we assumed the absence of any site-specific information.
	\item The inversion of each parameterization should be repeated multiple times (i.e., Ntrials $\geq$ 3) to check the inversion's convergence. Each trial should utilize at least 50,000 neighborhood algorithm models (It*Ns), although this may need to be increased if the parameterization is more complex than the examples considered in this work. We recommend Nr $\approx$100 as it showed the fastest convergence without the undo encouragement of outliers. The value of Ns0 was found to be unimportant, however we recommend it be no smaller than Nr.
	\item Parameterizations which significantly under-perform their counterparts should be rejected. We propose that a parameterization may be rejected if it meets the two following criteria: (1) the minimum misfit attained is significantly greater than those from other parameterizations, and (2) the parametrization contains either the largest or smallest number of layers (i.e., is at the edge in terms of parameterization simplicity/complexity) and has a Vs behavior that is inconsistent with its counterparts.
	\item At a minimum the lowest misfit model from multiple parameterizations should be reported to quantify the inter-parameterization uncertainty. Furthermore, the inter- and intra-parameterization uncertainty should be reported qualitatively, such as with a plot of 100 lowest misfit Vs profiles for each parameterization, like that in Figure \ref{fig:highvar}c and 11c, and quantitatively, such as with a plot of $\sigma_{ln,Vs}$ , like that in Figure \ref{fig:highvar}d and 11d, to communicate to the end user the relative (un)certainty of the inversion results.
\end{itemize}

\section{Acknowledgements}

The velocity models, experimental dispersion data, and inversion results used in the validation of the SWinvert workflow have been archived and made publically available through the DesignSafe-CI \citep{vantassel_surface_2020}. The authors hope that as others develop new and improved inversion algorithms the models considered here can be used as benchmarks to objectively and, with the proposed Vs misfit, quantitatively measure improvement. A Python package, \emph{SWprepost}, for surface wave inversion pre- and post-processing developed alongside this work has been released open-source \citep{vantassel_jpvantasselswprepost_2020}.  All inversion analyses were performed on the Texas Advanced Computing Center (TACC) resource Stampede2 using an allocation provided through the DesignSafe-CI \citep{rathje_designsafe_2017}. All inversion analyses presented in this work used the Dinver module of the open-source software Geopsy \citep{wathelet_geopsy_2020}. Geopsy is freely available for download at www.geopsy.org. To alleviate some of the computational expense of performing rigorous surface wave inversions an application on the DesignSafe-CyberInfracture, \emph{SWbatch}, has been released open-source \citep{vantassel_jpvantasselswbatch_2020}. \emph{SWbatch} allows the user to access high-performance computing resources through an intuitive and easy-to-use web interface to rapidly and conveniently invert experimental dispersion data considering multiple inversion parameterizations to address the problem's non-uniqueness and multiple trials per parameterization to address the problem's nonlinearity. The figures in this work were prepared using Matplotlib 3.1.2 \citep{hunter_matplotlib_2007} and Inkscape 0.92.4.

\bibliographystyle{plainnat}
\bibliography{swinvert}

\end{document}